\documentstyle[11pt,newpasp,twoside,epsf,psfig]{article}
\def\edcomment#1{\iffalse\marginpar{\raggedright\sl#1\/}\else\relax\fi}
\reversemarginpar

\begin{document}
\title{Multiplicity, kinematics and rotation rates of very young brown dwarfs
in Cha\,I}
\author{Viki Joergens, Ralph Neuh\"auser}
\affil{Max-Planck-Institut f\"ur Extraterrestrische Physik, 
	Giessenbachstr. 1, D-85748 Garching, Germany}
\author{Eike W. Guenther}
\affil{Th\"uringer Landessternwarte Tautenburg,
	Sternwarte 5, D-07778 Tautenburg, Germany}
\author{Matilde Fern\'andez}
\affil{Instituto de Astrof\'{\i}sica de Andaluc\'{\i}a (CSIC),
       Apdo. 3004, E-18080 Granada, Spain}
\author{Fernando Comer\'on}
\affil{European Southern Observatory,
	Karl-Schwarzschild-Str. 2, D-85748 Garching, Germany}

\begin{abstract}
We have studied twelve very young (1--5\,Myr)
bona fide and candidate brown dwarfs in the Cha\,I star forming region
in terms of their kinematic properties,
the occurrence of multiple systems among them as well as their rotational
characteristics.
Based on high-resolution spectra taken with UVES at the VLT (8.2\,m), radial and 
rotational velocities have been measured. 
A kinematic study of the sample showed that their radial velocity dispersion
is relatively small suggesting that they are not ejected during their
formation as proposed in recent formation scenarios.
By means of time-resolved UVES spectra, a radial velocity survey for close companions
to the targets was conducted.
The radial velocities of the targets turned out to be rather constant setting upper 
limits for the mass M$_2 \sin i$ of possible companions to 
0.1\,M$_{\mathrm{Jup}}$ -- 2\,M$_{\mathrm{Jup}}$. 
These findings hint at a rather low ($\leq$\,10\%)
multiplicity fraction of the studied brown dwarfs.
Furthermore, a photometric monitoring campaign of the targets
yielded the determination of rotational periods for 
three brown dwarf candidates in the range 
of 2.2 to 3.4 days. These are the first rotational periods for very young 
brown dwarfs and among the first for brown dwarfs at all.
\end{abstract}

\section{Introduction}

Although a large number of brown dwarfs have been detected up to date, 
fundamental parameters have been measured only for a small subset.
Consequently, there are still a lot of open questions, for example,
the mechanism that leads to the formation of brown dwarfs 
is still poorly constrained.

We have studied a population of twelve very young bona fide and candidate 
brown dwarfs in the Cha\,I star forming region, Cha\,H$\alpha$\,1 to 12 
(Comer\'on et al. 1999, 2000, Neuh\"auser \& Comer\'on 1998, 1999) 
by means of high-resolution spectroscopy ($\rm \lambda / \Delta \lambda=40\,000$)
with UVES at the 8.2\,m Kueyen telescope of the VLT
as well as by a photometric monitoring campaign carried out at the 
Danish 1.5\,m telescope at ESO.
These observations yielded radial and rotational velocities as well as rotational 
periods.
The studied brown dwarfs are only a few million years old and their observation 
allows insights into the formation and early evolution of brown dwarfs.

\section{UVES spectra I: Kinematic study}

\begin{figure}
\psfig{figure=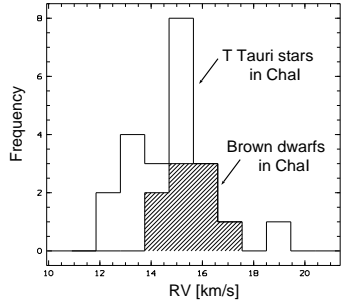,width=6cm,angle=270}
\caption{
Radial velocity distribution for 9 bona fide
and candidate brown dwarfs (hashed) and for 22 T~Tauri stars in Cha\,I.}
\end{figure}

High-resolution spectroscopic observations were conducted with the echelle 
spectrograph UVES attached to the 8.2\,m Kueyen telescope of the VLT.
At least two spectra separated by a few weeks have been taken 
of nine of the twelve targets.
The wavelength regime from 6600\,{\AA} to 10400\,{\AA} was covered
with a spectral resolution of $\rm \lambda / \Delta \lambda=40\,000$. 
Radial velocities have been measured by a cross-correlation of plenty of 
stellar lines of the object spectra against a template spectrum.
Telluric lines have been used as wavelength reference.
The achieved precision of the relative radial velocities range
between 80\,m\,s$^{-1}$ and 
600\,m\,s$^{-1}$, depending on the S/N of the individual spectra.

Recently, it has been proposed, that 
brown dwarfs are ejected protostars which have been cut off from the gas 
reservoir in the early accretion phase of the star formation process and could have 
therefore not accreted to stellar masses (Reipurth \& Clarke 2001).
A kinematic study of the sample (Joergens \& Guenther 2001) 
showed that their radial velocity dispersion
is relatively small (2.2\,km\,s$^{-1}$). It is significantly smaller than
the radial velocity dispersion of the T~Tauri stars in the field 
(3.6\,km\,s$^{-1}$) and slightly larger than that one of the surrounding 
molecular gas (1.2\,km\,s$^{-1}$). 
The distributions are displayed in Fig.\,1.
These results give suggestive evidence that there 
is no run-away brown dwarf among the studied sample.
It cannot be ruled out that some of them have a larger 3D velocity dispersion. 
However, the studied brown dwarfs occupy a field of less than
12$^{\prime}$\,$\times$\,12$^{\prime}$ at a distance of 160\,pc. 
Brown dwarfs born within this field
and ejected during the early accretion phase in directions with
a significant fraction perpendicular to the line-of-sight, would have flown out 
of the field a long time ago, given the age of the objects of 1--5\,Myr. 
This is true even for the smallest ejection velocities of 2\,km\,s$^{-1}$ 
calculated by theorists (see also next paragraph).
Therefore, the measurement of radial velocities is the very method to 
test if objects born in the field have significantly high velocities due to
dynamical interactions during their formation process. 

Very recent dynamical calculations (Bate et al. 2002, Sterzik \& Durisen 
this volume)
hint at rather small ejection velocities suggesting the possibility
that the imprint of the ejection in the kinematics might not be an observable 
effect. They predict a velocity dispersion of ejected brown dwarfs of 
2\,km\,s$^{-1}$ in 3D and 1.2\,km\,s$^{-1}$ in 1D.
On the other hand,
they predict that 10\% of the brown dwarfs have a larger velocity than 
5\,km\,s$^{-1}$ due to dynamical interactions.
The velocities of the bona fide and candidate brown dwarfs in Cha\,I 
cover a total range of only 2.6\,km\,s$^{-1}$. Therefore, it is
concluded that the ejection-model for the formation of brown dwarfs is
not a likely formation mechanism for the brown dwarfs in Cha\,I. 

\section{UVES spectra II: RV Survey for companions}

\begin{figure}
\plottwo{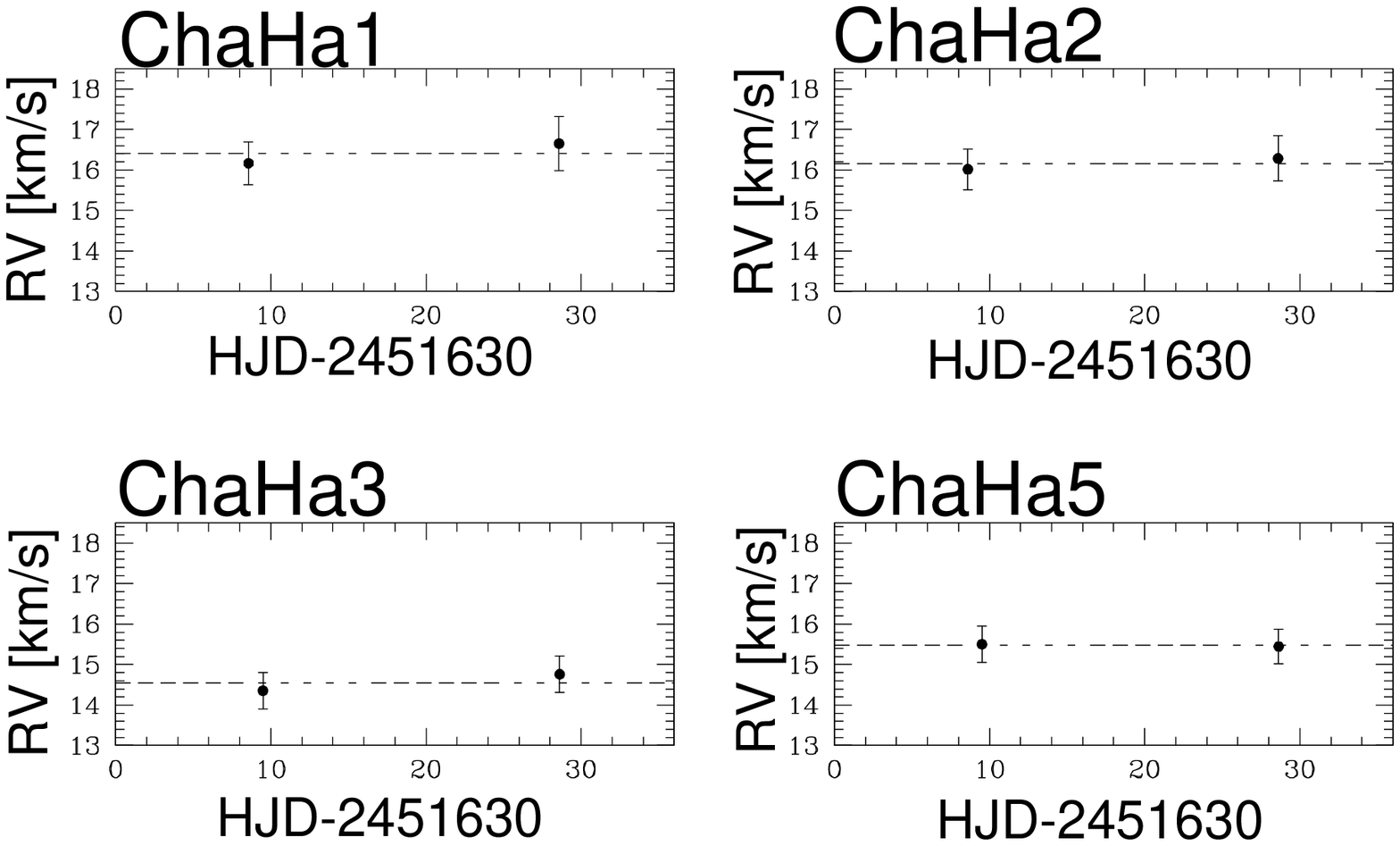}{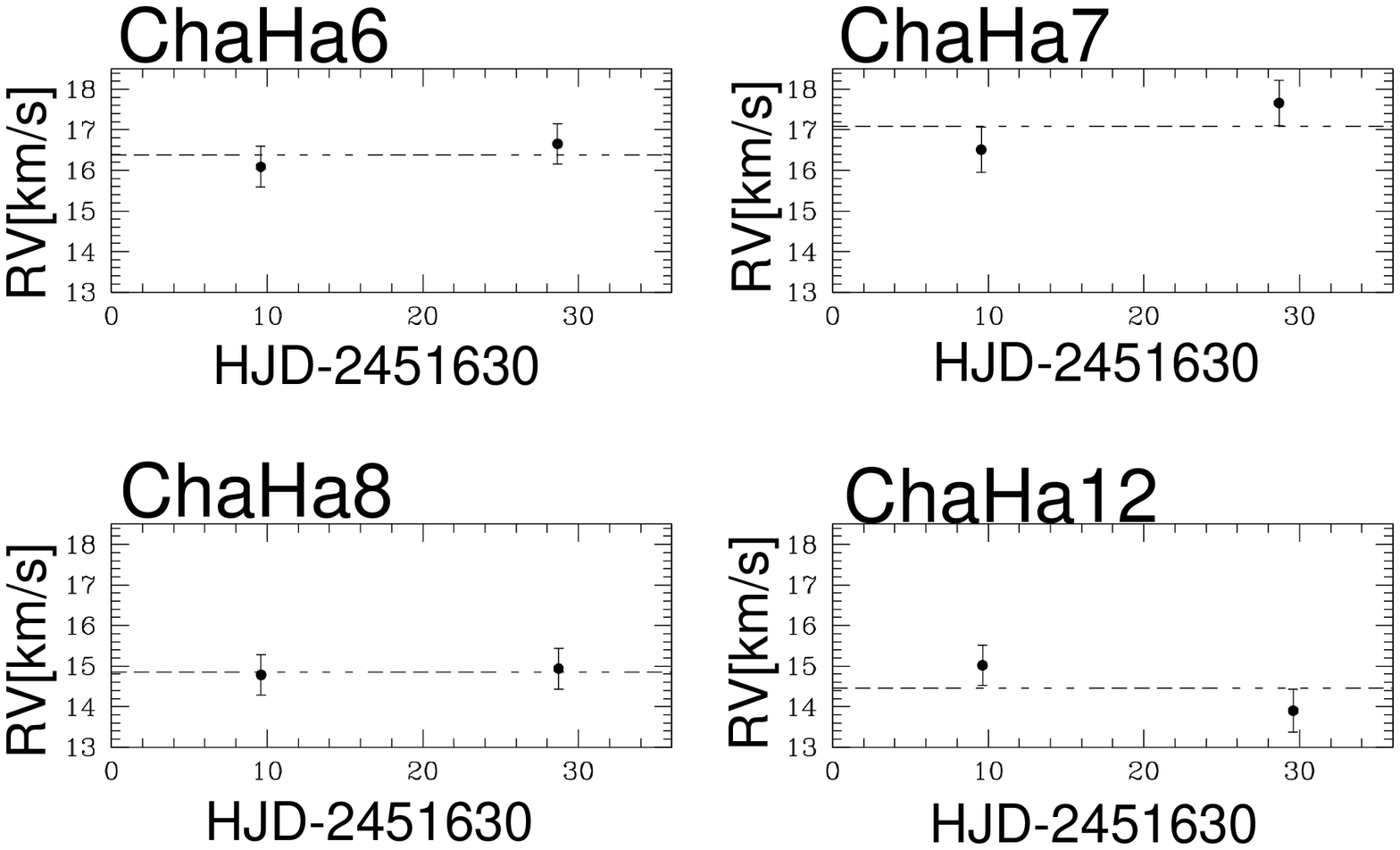}
\caption{Radial velocity vs. time.}
\end{figure}

\begin{figure}
\psfig{figure=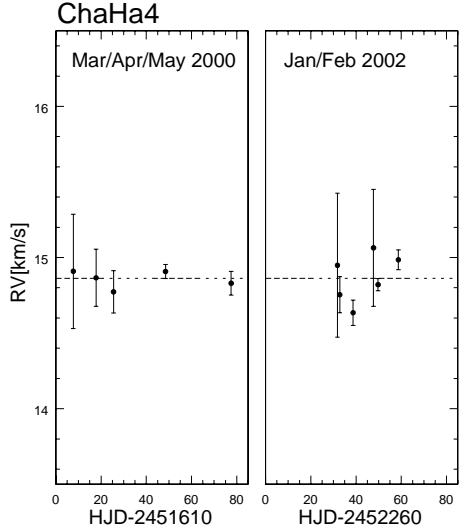,width=6cm}
\caption{Radial velocity vs. time.}
\end{figure}

By means of time-resolved UVES spectra, a radial velocity survey for close companions
to the targets was conducted.
For each of the objects Cha\,H$\alpha$\,1, 2, 3, 5, 6, 7, 8 and 12 
spectra have been taken in two nights separated by a few weeks.
For Cha\,H$\alpha$\,4 spectra have been taken in 11 nights, separated by weeks to years.

We have found that the radial velocities for Cha\,H$\alpha$\,1, 2, 3, 5, 6, 7, 8 and 12
are constant within the measurements uncertainties, whereas
Cha\,H$\alpha$\,4 shows small but significant variations (Fig.\,2).
Details will be published in a forthcoming paper 
(Joergens et al., in prep.).

An upper limit for the semiamplitude of a variation caused by a hypothetical
companion was determined for each object by assuming that the total
variability amplitude was recorded. 
Subsequently, upper mass limits for the hypothetical companions
have been estimated.
For this calculation an orbital separation of 0.1\,AU was assumed.
Primary masses from Comer\'on et al. (1999, 2000) have been used.
Furthermore, circular orbits and an inclination of 90$^{\circ}$ 
were assumed.
The assumption of $i=90^{\circ}$ means that the derived mass limits for companions
are upper limits for the minimum masses M$_2 \sin i$.

The upper limits for the semiamplitude of the 'constant' objects
(Fig.\,2, left panel) range between 30\,m\,s$^{-1}$
for Cha\,H$\alpha$\,5 and 600\,m\,s$^{-1}$ for Cha\,H$\alpha$\,7 and 12.
The deduced upper limits for 
M$_2 \sin i$ range between a tenth of a Jupiter mass (Cha\,H$\alpha$\,5)
and 1--2 Jupiter masses (Cha\,H$\alpha$\,7 and 12).
The small but significant variations of the radial velocity of Cha\,H$\alpha$\,4, 
if caused by a companion, correspond to a companion mass M$_2 \sin i$ of 
0.8\,M$_{\mathrm{Jup}}$.

In summary, no evidence for substellar companions around 
the targets has been found, aside from variations of the radial velocity of 
Cha\,H$\alpha$\,4, which could be due to a companion of less than 1\,M$_{\mathrm{Jup}}$.
There is the possibility that there are planetary mass companions with
masses below one Jupiter mass since the presented observations
were not sensitive enough to detect those. Furthermore, spectroscopic binaries 
in the sample might not have been detected due to 
the non-observation at the critical orbital phases.
However, the overall picture that arises is that the multiplicity fraction of the  
studied brown dwarfs is rather low ($\leq$ 10\%).

\section{Photometric monitoring: Rotational periods}

\begin{figure}
\plottwo{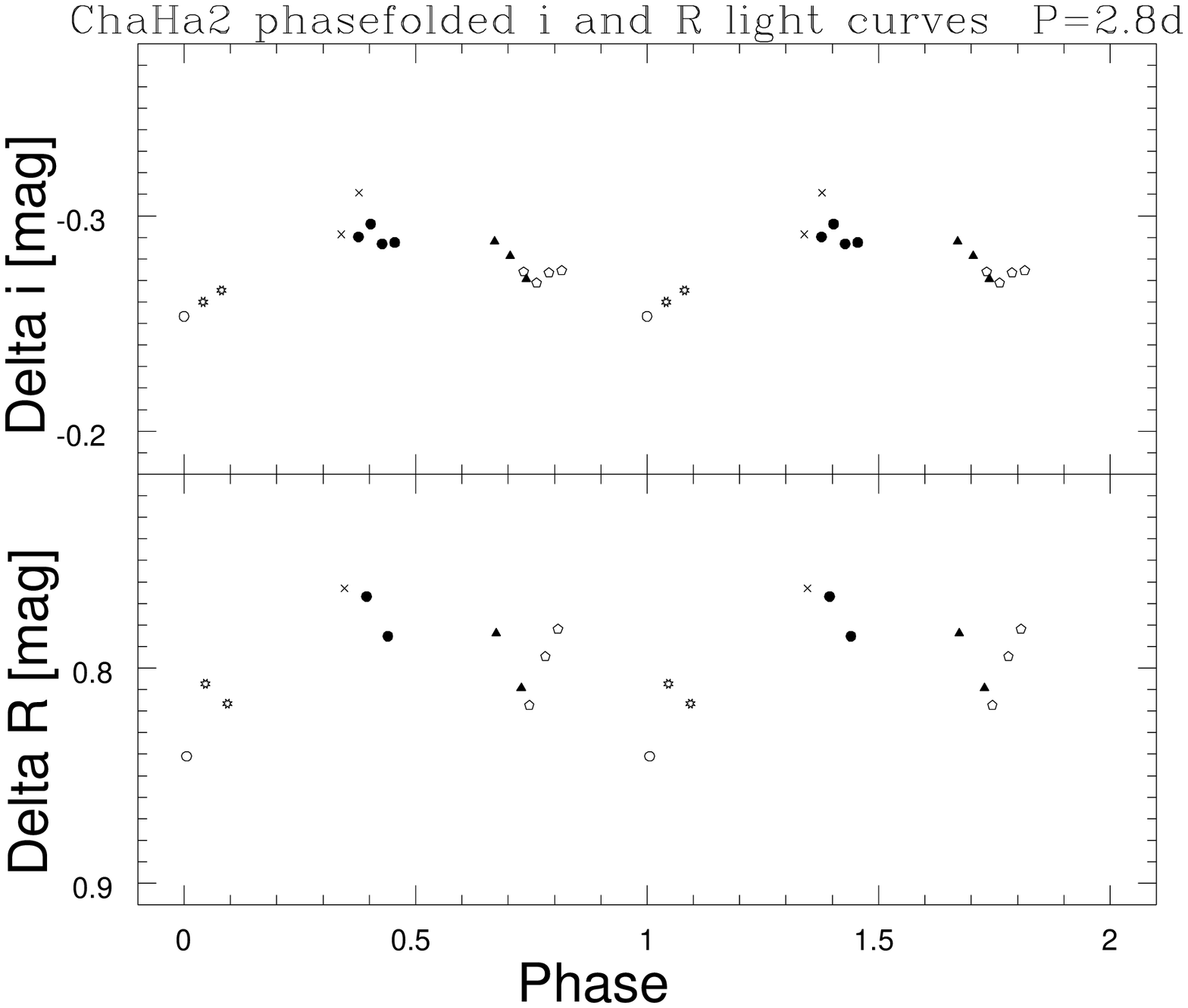}{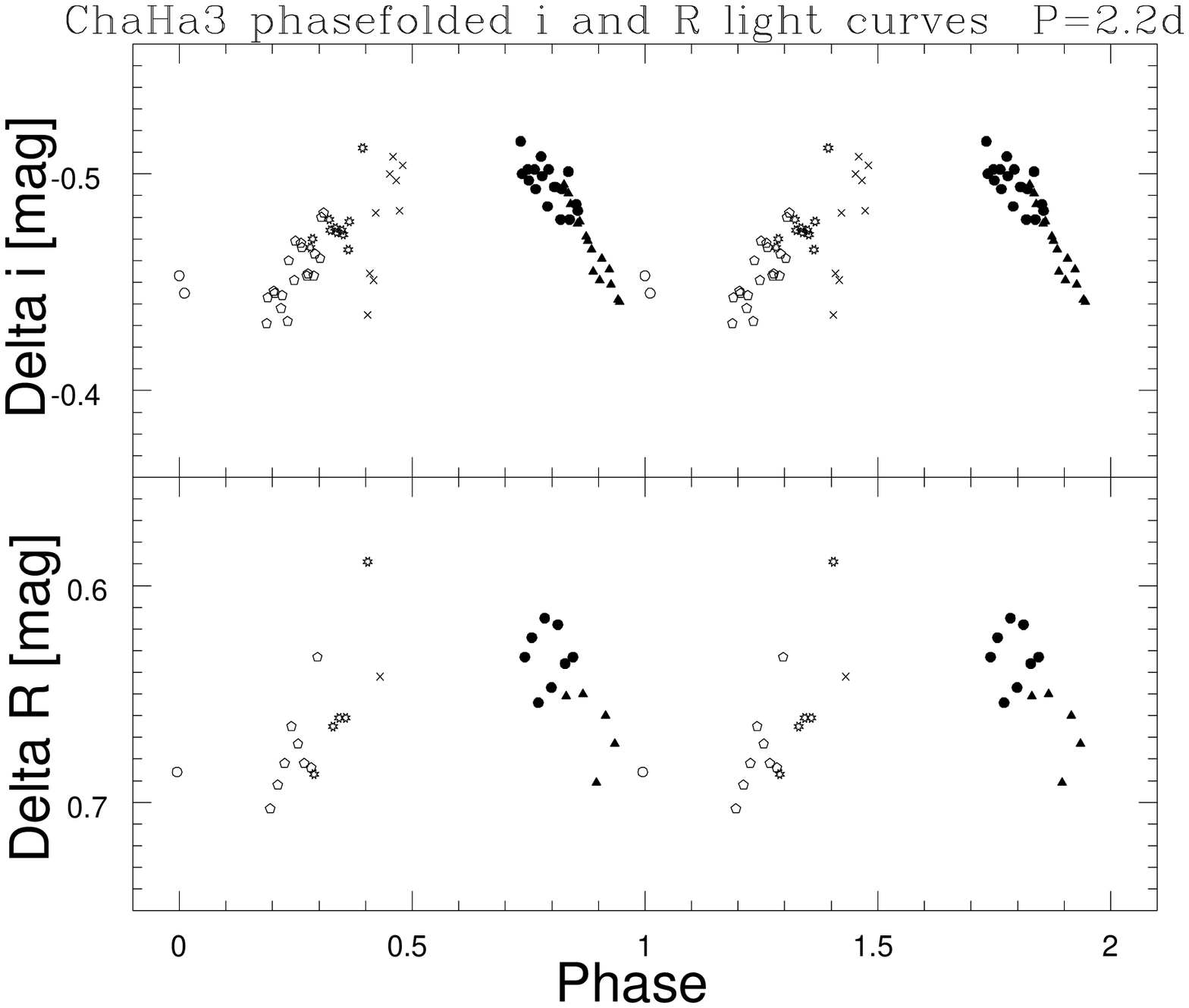}
\caption{Phase-folded i and R band light curves of 
Cha\,H$\alpha$\,2 and Cha\,H$\alpha$\,3. Different symbols 
denote data points obtained in different nights.}
\end{figure}

\begin{figure}
\plottwo{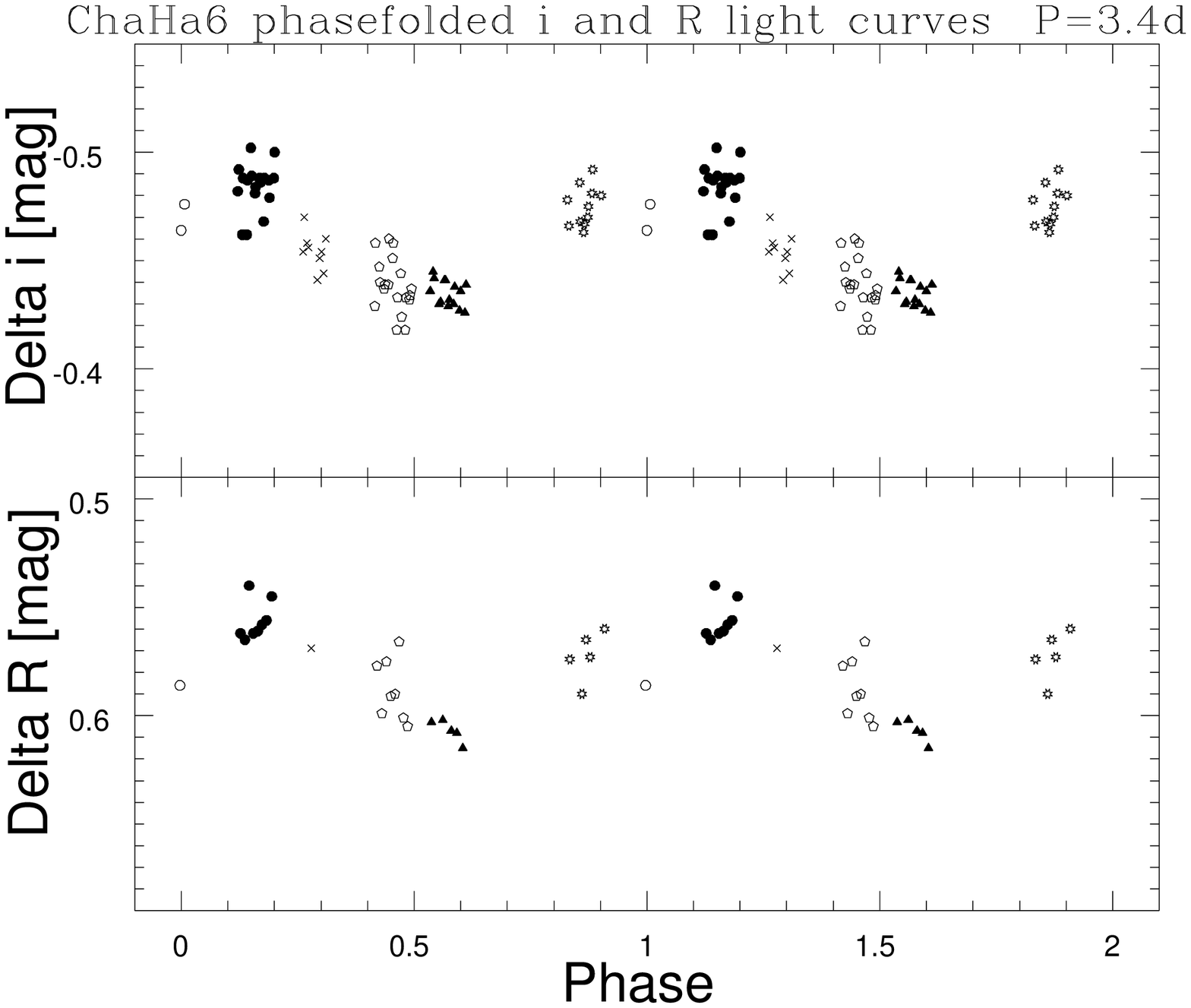}{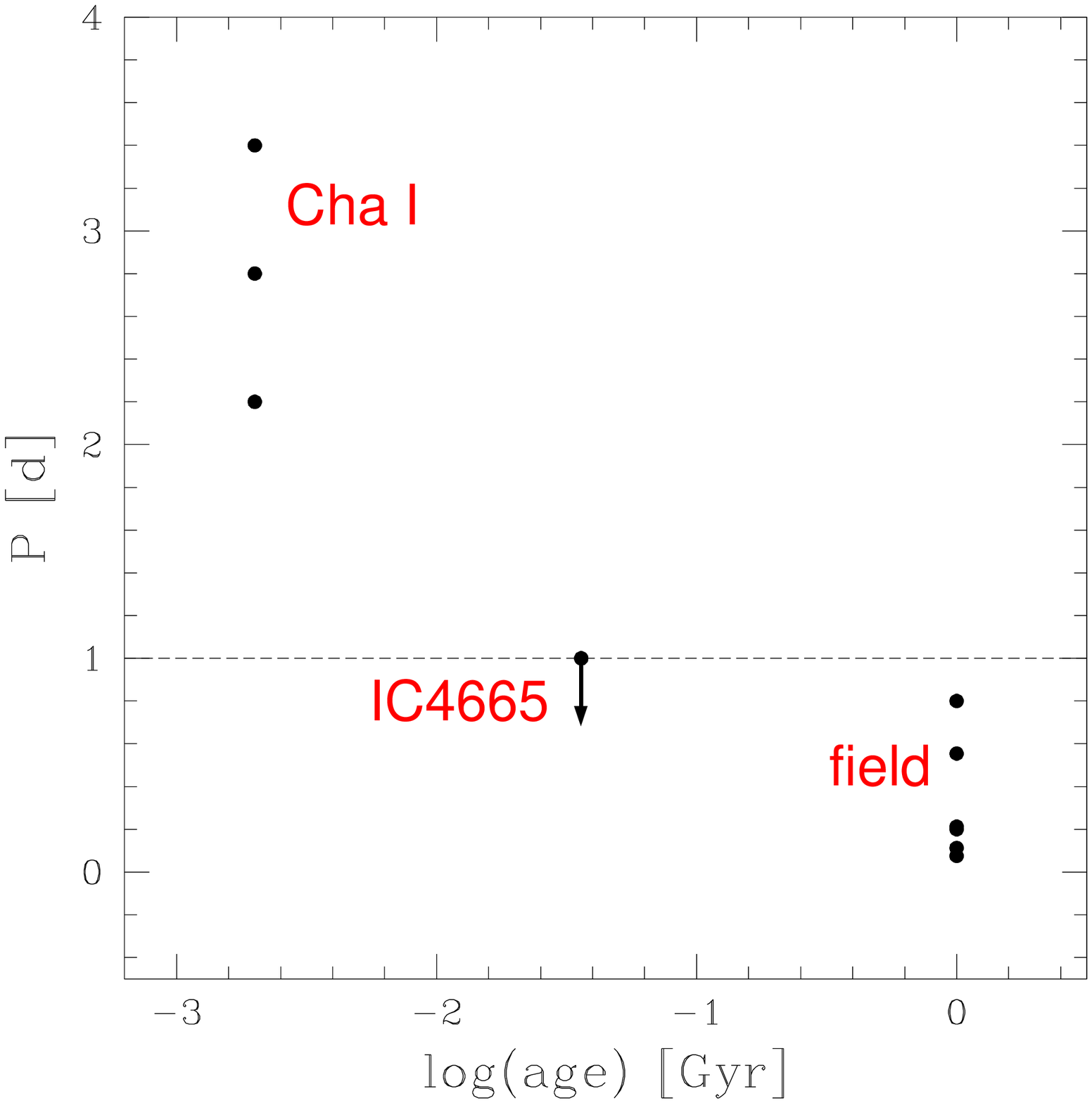}
\caption{Left: Phase-folded light curve of Cha\,H$\alpha$\,6.
Right: Evolution of angular momentum from 1\,Myr to 1\,Gyr.
Rotational period P vs. age for Cha\,H$\alpha$\,2, 3 and 6,
brown dwarfs in the cluster IC4665 (Eisl\"offel \& Scholz 2001)
and brown dwarfs in the field (Bailer-Jones \& Mundt 2001, Mart\'\i n et al. 2001,
Clarke et al. 2002).}
\end{figure}

A photometric monitoring campaign has been carried out at the Danish 1.5\,m telescope
at ESO in order to study the time dependence of the 
brightness of the objects. Rotational periods have been measured 
by tracing modulations of the light curves due to magnetically induced surface spots.
Rotational periods have been determined for the three brown dwarf candidates
Cha\,H$\alpha$\,2, Cha\,H$\alpha$\,3 and Cha\,H$\alpha$\,6
of 2.8\,d, 2.2\,d and 3.4\,d, respectively 
(Joergens et al. 2002a,b). 
See Fig.\,3 and 4 for phase-folded light curves.
The periods are consistent with measurements of
rotational velocities $v \sin i$ from UVES spectra for the objects 
(8--26\,km\,s$^{-1}$).
The observations show that brown dwarfs at an age of 1--5\,Myr 
display surface spots like T~Tauri stars and are moderately fast rotators
in contrast to rapidly rotating old brown dwarfs
consistent with them being in an early contracting stage.
A comparison with rotational periods 
at 36\,Myr (Eisl\"offel \& Scholz 2001) 
indicates that most of the acceleration of 
brown dwarfs takes place in the first 30 million years or less of their lifetime
(cf. Fig.\,4, right panel).

\section{Discussion and conclusions}

The kinematic study of brown dwarfs in Cha\,I based on the measurement of precise 
radial velocities from high-resolution UVES spectra showed that their radial velocities
have a dispersion of only 2.2\,km\,s$^{-1}$ and span a total range of only 
2.6\,km\,s$^{-1}$. We therefore conclude that the ejection-model for the formation 
of brown dwarfs (Reipurth \& Clarke 2001) is not a likely formation mechanism for 
the studied brown dwarfs since first calculations 
(Bate et al. 2002, Sterzik \& Durisen 2002, this volume) predict that 10\% of the 
brown dwarfs should have a larger velocity than 5\,km\,s$^{-1}$ due to 
the ejection. 

The small binary fraction ($\leq$ 10\%) found in the presented RV survey
is in agreement with the result of a direct imaging survey for wide, 
low-mass companions to the same objects
(Neuh\"auser et al. (2002), Joergens et al. (2001),
Neuh\"auser et al. this volume).
Combining both surveys, a wide range of possible companion 
separations has been covered. 
The exact separations depend on the companion masses. 
For example, for brown dwarf companions ($<$ 13\,M$_{\mathrm{Jup}}$)
to the targets, separations $<$3\,AU and between 50 and 1000\,AU were covered.
With more restricted separations ($<$0.1\,AU and 300--1000\,AU) the surveys
were sensitive also to companion masses down to 1\,M$_{\mathrm{Jup}}$.

These results seem to be in contrast with the high multiplicity fraction observed for 
very young stars and may suggest that brown dwarfs form not by direct
collapse of unstable cloud cores as stars.
However, a significant comparison of the multiplicity fraction of stars and brown 
dwarfs at very young ages is still hampered by small-number statistics in the 
substellar regime. Such a study has to compare multiplicity fractions in a certain
separation range, which has to agree for both samples.
Furthermore, it is possible that the 
gravitational collapse of cloud cores of brown dwarf masses
does not yield as much multiple systems as
the collapse of clouds of solar masses.
In addition, stars can have companions, which have
only a tenth of the mass of the primary, whereas a companion of a brown dwarf with
such a low mass ratio would be already a planet. 
However, there is still a lot to do for observers as well as for theorists
in order to understand in which way 
objects of brown dwarf masses are formed.

Finally, we have determined rotational periods for the three brown dwarf candidates 
Cha\,H$\alpha$\,2, Cha\,H$\alpha$\,3 and Cha\,H$\alpha$\,6 of 2.2 to 3.4 days
by means of a photometric monitoring campaign.
These are the first rotational periods for very young brown dwarfs and among the
first for brown dwarfs at all. They provide valuable data 
points in an as yet almost unexplored region of the substellar period-age diagram.
A comparison of the determined rotational periods
at the age of a few million years with rotational properties
of older brown dwarfs ($>$36\,Myr, Eisl\"offel \& Scholz 2001)
shows that most of the acceleration of brown dwarfs
during the contraction phase takes place within the first 30\,Myr or less.
It is known that Cha\,H$\alpha$\,2 and 6 have optically thick disks 
(Comer\'on et al. 2000), therefore magnetic braking due to 
interactions with the disk may play a role for them.
This is suggested by the fact, that among 
the three brown dwarf candidates with determined periods, the one without
a detected disk, Cha\,H$\alpha$\,3, has the shortest period.
If the interaction with the disk is responsible for the braking, 
the results from Eisl\"offel \& Scholz (2001) indicate that 
brown dwarf inner disks have been dissipated at an age of 36\,Myr.
These limits for the time scale of disk dissipation
for brown dwarfs are very similar to those for T~Tauri stars, which are
known to dissolve their inner disks within about the first 10\,Myr 
(e.g. Calvet et al. 2000). 

Further measurements of rotational periods for brown dwarfs are much-needed 
in order to complete the picture of angular momentum evolution in the substellar regime
as well as that of rotationally induced phenomena, like dynamo activity and meteorological processes.


\end{document}